\documentclass[12pt]{article}
\usepackage{amsmath}
\usepackage{graphicx}
\usepackage{enumerate}
\usepackage{natbib}
\usepackage{url} 

\usepackage{color}
\usepackage{amssymb}
\usepackage{amsmath}
\usepackage{listings}
\usepackage{tikz}
\usetikzlibrary{arrows,positioning,decorations.pathreplacing,decorations.pathmorphing,calc}
\usepackage{authblk} 
\usepackage{bm}

\newcommand{\blind}{0}

\newcommand{\argmin}{\operatornamewithlimits{argmin}}

\newcommand{\BetaDist}{\operatorname{Beta}}

\newcommand{\X}{\bm{X}}

\definecolor{res}{RGB}{87,144,34}
\definecolor{sen}{RGB}{160,38,97}
\definecolor{green2}{RGB}{51,140,23}
\definecolor{grey}{RGB}{100,100,100}
\definecolor{greylight}{RGB}{160,160,160}
\definecolor{greydark}{RGB}{60,60,60}

\begin{document}

\def\spacingset#1{\renewcommand{\baselinestretch}%
{#1}\small\normalsize} \spacingset{1}


\title{
  Leveraging Historical Data for
  High-Dimensional Regression Adjustment, a
  Composite Covariate Approach}
\author[1]{Samuel Branders}
\author[1]{Alvaro Pereira}
\author[1]{Guillaume Bernard}
\author[2]{Marie Ernst}
\author[2,3]{Adelin Albert}
\affil[1]{\small Tools4Patients s.a, 11 rue Granbonpr\'e, Bte 9, 1435 Mont-Saint-Guibert, Belgium}
\affil[2]{\small Biostatistics, Medico-economic information Department, University Hospital of Li\`ege, 4000 Li\`ege, Belgium}
\affil[3]{\small Department of Public Health, University of Li\`ege, 4000 Li\`ege, Belgium}

\if0\blind
{
  \maketitle
} \fi

\if1\blind
{
  \bigskip
  \bigskip
  \bigskip
  \begin{center}
      {\LARGE\bf Leveraging historical data for
        high-dimensional regression adjustment, a
        composite covariate approach}
  \end{center}
  \medskip
} \fi

\bigskip
\begin{abstract}
\noindent The amount of data collected from patients involved in clinical trials is continuously growing.
All patient characteristics are potential covariates that could be used to improve clinical trial analysis and power.
However, the restricted number of patients in phases I and II studies limits the possible number of covariates included in the analyses.
In this paper, we investigate the cost/benefit ratio of including covariates in the analysis of clinical trials.
Within this context, we address the long-running question ``What is the optimum number of covariates to include in a clinical trial?''
To further improve the cost/benefit ratio of covariates, historical data can be leveraged
to pre-specify the covariate weights,
which can be viewed as the definition of a new composite covariate.
We analyze the use of a composite covariate while estimating the treatment effect in small clinical trials.
A composite covariate limits the loss of degrees of freedom and the risk of overfitting.

\end{abstract}

\noindent%
\vfill

\newpage
\spacingset{1.5} 


\section{Introduction}
\label{sec:intro}



The amount of information collected from patients involved in clinical trials is steadily growing, in particular with the advent of genomics and proteomics.
All collected baseline characteristics are potential covariates linked with the patient's outcome.
Some covariates may not always be of primary interest in a randomized clinical trial (RCT), however they could be used to explain the variability of the patient's response and improve the study power when assessing treatment efficacy.

\bigskip
\noindent The adjustment for baseline covariates to improve the efficiency of randomized clinical trial analysis can be done in many ways. For example, patients may be stratified according to various factors before randomization. Age, gender, social status, or race are commonly used stratification criteria that may interact with outcome.
One of the most traditional methods however is to include covariates in a general regression equation of the form $Y = \mu + \gamma Z + \beta^T X + \varepsilon$,
where $Y$ is the outcome variable, $\mu$ a constant, $Z$ the treatment, $X$ the vector of covariates and $\varepsilon$ the error term. The parameter $\gamma$ measures the adjusted treatment effect and $\beta$ is the vector of regression coefficients of covariates.

\bigskip
Including covariates associated with the study outcome could greatly improve the efficiency and power of the trial.
They could correct for potential bias coming from baseline covariate imbalance between the study arms.
However, adding covariates in the analysis comes with a cost in degrees of freedom.
As such, regression adjustment should be seen as a trade-off between explained variance and loss of degrees of freedom.
Clearly, for small trials, the number of covariates to be included in the model must be limited.
There are many rules-of-thumb on the number of covariates that can be included in an analysis~(\cite{Austin2015,Schmidt1971}).
A common one is to have 10 subjects per variable in the model.
The problem with this heuristic rule and other approaches is the variance explained by the covariates that is not taken into account.

\bigskip
In this paper, we intend to investigate the cost/benefit ratio of including covariates in the analysis of RCTs.
Instead of focusing on the estimation of the treatment effect, we search to minimize its sampling variance, i.e., to increase its statistical precision, while considering the covariates as nuisance factors.
Within this context, we address the long-running question ``What is the optimum number of covariates to include in a clinical trial?''

\bigskip
To improve the cost/benefit ratio of covariates, their weights in the model could be estimated from historical data accumulated from outside sources.
In cancer research, the treatment effect is often adjusted for a single baseline prognostic index.
For example, the Nottingham prognostic index (NPI) defined by \cite{Galea1992} and used in breast cancer incorporates the size and grade of the tumor as well as the nodal status.
Adjusting for all three parameters would explain more variance but at an additional cost in terms of degrees of freedom (\cite{Moons2009,INHLPFP93,Galea1992,Keys1972}).
There are numerous examples of this kind.
We investigate the benefits of replacing individual covariates by a composite covariate fitted on external data,
motivated by the recent advances in placebo effect characterization~(\cite{Horing2014,Pereira2016,Vachon-presseau2018}).
Indeed, the multiple facets and high-dimensionality of the placebo effect makes any adjustment difficult and could therefore advantageously benefit from a composite covariate approach.


\bigskip
This work is structured as follows. In Section \ref{sec:model}, we introduce the general model describing the relationship between the patient's outcome and the treatment while accounting for a vector of potential covariates. It serves as the generative data model in our theoretical developments, simulations and illustrations.
Section \ref{sec:variance} focuses on the sampling variance of the treatment effect with and without covariates adjustment.
In Section \ref{sec:optimum}, we propose an approach to select covariates minimizing the expected treatment effect sampling variance based on historical data.
In Section \ref{sec:composite}, we discuss the relative efficiency of combining covariates a priori as a way to limit the number of parameters to be fitted in the model.
In Section \ref{sec:simulation}, we perform simulation studies to demonstrate the benefit of the composite covariate approach.
We conclude with a brief discussion section.





\section{Generative model of the data}
\label{sec:model}

Suppose we focus on the treatment effect only. Then, the response model writes
\begin{equation}\label{eq:model0}
Y = \mu + \gamma Z + U
\end{equation}
where for simplicity $Z$ the treatment variable equals 1 for treatment and 0 for placebo and $U$ is the error term. The random variable $U\sim \mathcal{N}(0,\sigma^2_u)$ accounts for all factors not linked with treatment.
Since in most clinical studies, patients are randomized between arms,
the independence between $Z$ and $U$ can be assumed.

\bigskip
The random variable $U$ may in turn be expressed as a linear function of the covariates $X$, namely
\begin{equation}\label{eq:modelU}
U = \beta^T X + \varepsilon
\end{equation}
where $X=(X_1,\ldots,X_p)^T$ is a vector of $p$ covariates and the error term $\varepsilon$ is assumed to be normally distributed $\mathcal{N}(0,\sigma^2_\varepsilon)$, independently of $X$.

Thus, by combining Equations \eqref{eq:model0} and \eqref{eq:modelU} and assuming $\mu=0$ without loss of generality, the general regression model writes
\begin{equation}\label{eq:modelp}
Y = \gamma Z + \beta^T X + \varepsilon.
\end{equation}

For the sake of simplicity, this paper mainly focuses on the two groups setting: placebo versus active.
However, all results can easily be generalized to $g$ study groups as presented in Appendix~\ref{sec:moreGroups}.


\section{Variance of the estimated treatment effect}
\label{sec:variance}

To estimate the added value of the covariates, we should compare the variance of the estimators of ${\gamma}$ with and without covariates.
To avoid any confusion, we denote by $\hat{\gamma_0}$ the ordinary least squares (OLS) estimator of $\gamma$ when no covariate is used in the regression (Equation \ref{eq:model0}) and by
$\hat{\gamma_p}$ when $p$ covariates are included in the regression (Equation \ref{eq:modelp}).

\bigskip
Now, consider a random sample of $n$ observations with responses ($i=1,\ldots,n$)
\begin{equation}\label{eq:model}
y_i = \gamma z_i + \beta_1  x_{i1} + \cdots + \beta_p  x_{ip}+ \varepsilon_i
\end{equation}
where $y_i$ is the response for patient $i$, $z_i$ is the treatment assigned, $x_{i1},\ldots,x_{ip}$ are the observed covariates, and $\varepsilon_i$ is the error term which is independent from $z_i$ and $x_{ij}$.
The vector of treatment assignment is denoted $z = (z_1, \dots, z_n)^T$.
The design matrix is denoted $\X=(x_{ij}-\bar{x}_j)_{1\leq i \leq n, 1\leq j \leq p}$.


\subsection{Without covariate ($p=0$)}

When no covariates are included in the model, Equation \eqref{eq:model} is simplified as
\begin{equation*}
y_i= \gamma z_i + u_i
\end{equation*}
for $i=1,\ldots,n$ and $u_i \stackrel{i.i.d.}{\sim} \mathcal{N}(0,\sigma_u^2)$,
the OLS estimated treatment effect writes
\begin{equation}\label{eq:gamma0}
\hat{\gamma}_0 = \frac{\sum_{i=1}^n (z_i-\bar{z})(y_i-\bar{y})}{\sum_{i=1}^n(z_i-\bar{z})^2}.
\end{equation}
and its sampling variance, conditional on $z$, is given by the expression
\begin{equation}\label{eq:var0}
\mbox{Var}(\hat{\gamma}_0 | z) = \frac{\sigma_u^2}{\sum_{i=1}^n(z_i-\bar{z})^2}.
\end{equation}
Observe that $\sigma_u^2$ can be estimated without bias by $\hat{\sigma}_u^2 = SSE_0/(n-2)$, where $SSE_0=\sum_{i=1}^n (y_i- \hat{\gamma}_0 z_i)^2 = \sum_{i=1}^n \hat{u}_i^2$ is the residual sum of squares with $(n-2)$ degrees of freedom.
Thus, the estimated sampling variability writes
\begin{align}
\widehat{\mbox{Var}}(\hat{\gamma}_0) &= \frac{\hat{\sigma}_u^2}{\sum_{i=1}^n(z_i-\bar{z})^2} \label{eq:gamma0b2}\\
&= \frac{\sum_{i=1}^n \hat{u}_i^2}{(n-2)\sum_{i=1}^n(z_i-\bar{z})^2} \label{eq:gamma0b}
\end{align}

\subsection{With $p$ covariates}
In a similar way, when $p$ covariates are included in the model, Equation~\eqref{eq:model} is used with
$\varepsilon_i\stackrel{i.i.d.}{\sim} \mathcal{N}(0,\sigma_\varepsilon^2)$.
The OLS estimator of $\gamma$ is given by the expression
\begin{equation}\label{eq:gammap}
\hat{\gamma}_p = \frac{\sum_{i=1}^n \hat{r}_i (y_i-\bar{y})}{\sum_{i=1}^n \hat{r}_i^2}.
\end{equation}
where $\hat{r}_i$ are the OLS residuals from the regression of $Z$ on the $p$ covariates $X_1, \dots, X_p$ based on $n$ observations $(z_i, x_{i1}, \dots, x_{ip})$, $i=1,\dots,n$.
The sampling variance of the OLS estimator, conditional on $z$ and $\X$, writes
\begin{equation}\label{eq:varp}
\mbox{Var}(\hat{\gamma}_p | z, \X) = \frac{\sigma^2_\varepsilon}{(1 - \hat{R}_{z:\X}^2) \sum_{i=1}^{n} (z_i - \overline{z})^2}
\end{equation}
where $\hat{R}_{z:\X}^2$ is the estimated multiple coefficient of determination of the regression of $Z$ on the $p$ covariates $X$.

As before, we note that $\sigma^2_\varepsilon$ can be estimated without bias by $\hat{\sigma}^2_\varepsilon = SSE_p/(n-p-2)$ where $SSE_p=\sum_{i=1}^n (y_i - \hat{\gamma}_p z_i - \hat{\beta}_1 x_{i1} \cdots - \hat{\beta}_p x_{ip})^2 = \sum_{i=1}^n \hat{\varepsilon}_i^2 $ is the residual sum of squares with $(n-p-2)$ degrees of freedom.
As such, the estimated sampling variability writes
\begin{align}
\widehat{\mbox{Var}}(\hat{\gamma}_p) &= \frac{\hat{\sigma}^2_\varepsilon}{(1 - \hat{R}_{z:\X}^2) \sum_{i=1}^n(z_i-\bar{z})^2} \label{eq:vargammap3}\\
&= \frac{\sum_{i=1}^n \hat{\varepsilon}_i^2}{(n-2-p) (1 -\hat{R}_{z:\X}^2) \sum_{i=1}^n(z_i-\bar{z})^2} \label{eq:vargammap}
\end{align}

\subsection{Benefits of including covariates}

Conditional on $z$ and $\X$, a gain in the statistical precision of the estimated treatment effect is obtained if
\begin{equation}
    \mbox{Var}(\hat{\gamma}_p) < \mbox{Var}(\hat{\gamma}_0).
\end{equation}
Using Equations \eqref{eq:var0} and \eqref{eq:varp}, the inequality writes
\begin{equation*}
    \frac{\sigma_\varepsilon^2}{(1-\hat{R}_{z:\X}^2)\sum_{i=1}^n(z_i-\bar{z}^2)} < \frac{\sigma_u^2}{\sum_{i=1}^n(z_i-\bar{z}^2)}
\end{equation*}
or
\begin{equation}
    \frac{1}{(1-\hat{R}_{z:\X}^2)} \frac{\sigma_\varepsilon^2}{\sigma_u^2} < 1.
\end{equation}
However, the actual values of the covariates $X$ are not known in advance and should be treated as random variables.
Therefore, the inequality should hold on average by taking the expectation over the joint distribution of $Z$ and $X$.
In doing so, we define the relative efficiency of including the $p$ covariates in the model, namely
\begin{align}\label{eq:RE}
    RE_p := E_{Z,X} \left[ \frac{1}{(1-\hat{R}_{z:\X}^2)} \frac{\sigma_\varepsilon^2}{\sigma_u^2} \right].
\end{align}
As discussed in \cite{Shieh2020}, it is constructive to assume the covariates have independent and identical normal distribution for each patient
\begin{align}
    X \sim \mathcal{N}_p(\theta, \bm{\Sigma})
\end{align}
where $\theta$ is a $p \times 1$ vector and $\bm{\Sigma}$ is a $p \times p$ positive-definite variance-covariance matrix.
With the independence between $Z$ and $X$, $\hat{R}_{z:\X}^2$ follows a beta distribution, $\BetaDist(p/2, (n - p - 1)/2)$.
Therefore, direct computation gives
\begin{align}
E\left[\frac{1}{ (1 - \hat{R}_{z:\X}^2)}\right]
&= \frac{(n - 3)}{(n - p - 3)}. \label{eq:expRzb}
\end{align}

Since the relative efficiency should be less than 1, by combining Equations \eqref{eq:RE} and \eqref{eq:expRzb}, we get the condition
\begin{equation}
RE_p = \frac{(n - 3)}{(n - p - 3)} \frac{\sigma_\varepsilon^2}{\sigma_u^2} < 1. 
\end{equation}
By setting
\begin{equation}\label{eq:defnup}
\nu_p := 1 - \frac{\sigma_\varepsilon^2}{\sigma_u^2},
\end{equation}
the relative residual variance of including $X$ in the model, namely the proportion of variance of $Y$ explained by the $p$ covariates after accounting for treatment, the condition can be written
\begin{equation}
RE_p = \frac{(n - 3)}{(n - p - 3)} (1-\nu_p) < 1. \label{eq:rep}
\end{equation}
As a consequence, the $p$ covariates included in the model improve the statistical precision of the estimator of $\gamma$ if
\begin{equation}\label{eq:gainnup}
\nu_p > \frac{p}{n - 3}.
\end{equation}
Equivalently, the number of variables to be included should satisfy the inequality
\begin{equation}
p < (n-3) \nu_p.
\end{equation}

These equations can be easily extended in a more general setting with more than two groups.
This extension to $g$ groups is presented in Appendix~\ref{sec:moreGroups}.
The result is a slightly modified version of Equation~\eqref{eq:rep}
\begin{equation}
    RE_p = \frac{(n - g - 1)}{(n - p - g - 1)} (1-\nu_p) < 1. \label{eq:repG}
\end{equation}
For the sake of simplicity, we focus here on the two groups setting.
However, the reader should keep in mind that $n-3$ and $n-p-3$ can be viewed as $n-g-1$ and $n-p-g-1$.


\section{Optimal number of covariates}
\label{sec:optimum}

In the previous section, we show how to estimate the maximum number of covariates.
The next obvious question is: ``What is $p$, the optimal number of covariates to be included in an analysis?''
To answer this question, hypotheses should be made on the gain in variance brought by each individual covariate.
In a simplistic approach, let's first assume that the $p$ covariates are independent and explain the same amount of variance, $\nu_1$.
Due to the independence assumption, the relative efficiency (Equation \eqref{eq:rep}) becomes:
\begin{equation}
    RE_p = \frac{(n - 3)}{(n - p - 3)} (1- p \nu_1)
\end{equation}
The relative efficiency is monotonically decreasing with $p$ if $\nu_1 > 1/(n-3)$ and increasing otherwise.
As such, depending on $\nu_1$, the optimal number of covariates that should be included in the analysis is either none or all of them.
This results has an easy and interesting practical application for the a priori selection of covariates.
Assuming the independence between the covariates, $RE_p$ can only decrease while including covariates explaining more than $1/(n-3)$.
This threshold can easily be checked on prior data while computing the correlation between each covariate and the outcome.
As such, one could include in the model all covariates with an expected correlation with the outcome above $1/\sqrt{n-3}$.




\bigskip
In a more realistic scenario, the amount of explained variance would not be equally spread amongst all covariates and the covariates might not be independent of each other.
In a clinical research context, it is relatively fair to assume that a ranking from the most to the least interesting covariates is known.
Then using Equation \eqref{eq:rep}, the optimal number of covariates is:
\begin{align}
    \argmin_p \frac{(n - 3)}{(n - p - 3)} (1-\nu_p)
\end{align}
with $p \in [0, n-4] \cap \mathbb{N}$.
Here, $\nu_p$ is the population coefficient of determination of the regression with respect to the $p$ most interesting covariates and could be estimated from previous data in the same indication.


Indeed, developing a new drug requires the conduct of several successive clinical trials.
In most cases, it is fair to assume the existence of previous study data in the same indication.
As the covariates are expected to be independent of the study treatment, previous data could come from studies investigating other compounds as well.
These previous study data could be leveraged to estimate all $\nu_p$ using the formula propose by \cite{Olkin1958}:
\begin{align}
    \hat{\nu_p} = 1 - \frac{(m - 3)}{(m - p - 1)} (1 - \hat{R}^2) F\left(1, 1, \frac{(m - p + 1)}{2}, 1 - \hat{R}^2 \right)
\end{align}
where $\hat{R}^2$ is the multiple R-squared of the regression of the $U$ by the covariates $X$,
$m$ the number of patients from the previous existing data, and
$F$ the hypergeometric function.
We used $m$ to make clear that the number of patients from the previous data is not the same as $n$ the number of patients from the current (or planned) study of interest.

Of note, the assumption that a ranking of the covariates is known is not strictly required.
Indeed, one could compute $\nu_p$ while testing all possible sets of $p$ covariates.
However, this might be computationally intensive and prone to overfitting.


\section{Composite covariate approach}
\label{sec:composite}

Assuming that historical data exist, as in the previous section, could we do better than only estimating the optimum number of covariates?
The main problem with the use of covariates is the associated loss in degrees of freedom.
To avoid this issue, we could derive the vector of covariates weights, $\beta$, directly from historical data.
More simply, we define a new covariate $W=f(X)$ as a composition of the $p$ individual covariates.
This composite covariate is then used as any covariate in the following studies while limiting the loss of degrees of freedom.
Specifically, the model given by Equation \eqref{eq:modelp} simplifies as follows
\begin{align}\label{eq:compositemodel}
Y &= \gamma Z + \beta W + \varepsilon.
\end{align}
We have a gain in statistical precision of the treatment effect (denoted $\hat{\gamma}_W$) if the relative efficiency of the new composite covariate, $W$, is less than the relative efficiency of using $X$, namely if,
\begin{align}
E_{Z,X} \left[ \frac{\mbox{Var}(\hat{\gamma}_W)}{\mbox{Var}(\hat{\gamma}_0)} \right]
&< E_{Z,X} \left[ \frac{\mbox{Var}(\hat{\gamma}_p)}{\mbox{Var}(\hat{\gamma}_0)} \right] 
\end{align}
or, using expression in Equation \eqref{eq:rep} for $p=1$ and general $p$, if
\begin{equation}
\frac{(n - 3)}{(n -4)} (1-\nu_W) < \frac{(n - 3)}{(n - p - 3)} (1-\nu_p)
\end{equation}
or
\begin{equation}
\frac{(n - p - 3)}{(n -4)} \frac{1-\nu_W}{1-\nu_p} < 1
\end{equation}
where $\nu_W$ is the relative proportion of the variance of $U$ explained by the composite covariate in the population. Thus, for a benefit of the composite covariate with respect to $p$ covariates, we need to have
\begin{equation}\label{eq:nuWnup}
\nu_W > 1 - \frac{n-4}{n-p-3}(1-\nu_p).
\end{equation}

\begin{figure}[ht]
    \center
    \begin{tikzpicture}

        \node (fig1) at (0,0) {\includegraphics[scale=0.60]{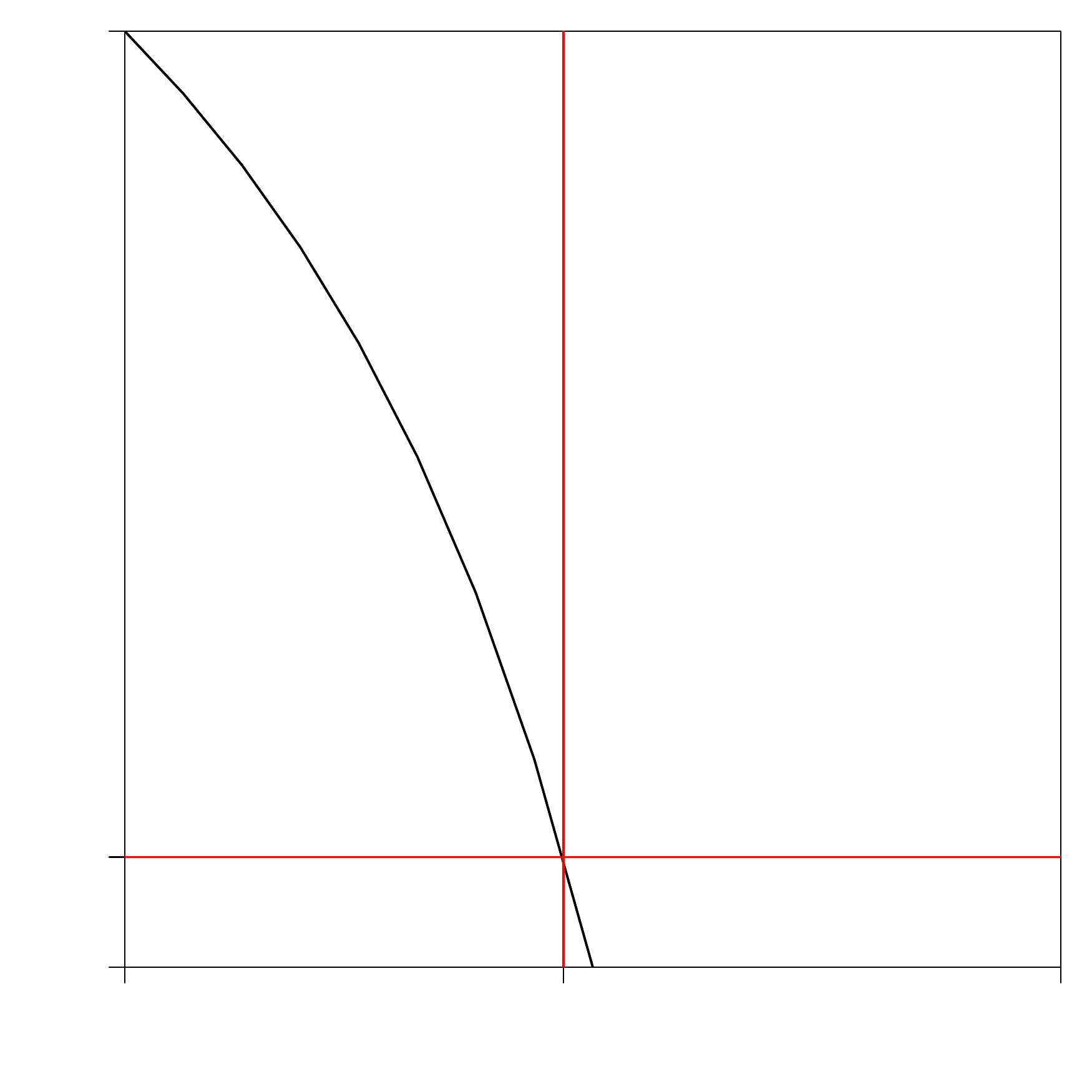}};

        \node[black] (s1) at (2.7,0.3) {$\hat{\gamma}_W$, $\hat{\gamma}_0$, $\hat{\gamma}_p$};
        \node[anchor=west,black] (s2) at (-2.4, 3.5) {$\hat{\gamma}_W$, $\hat{\gamma}_p$, $\hat{\gamma}_0$};
        \node[black] (s2) at (-2.55,-1.2) {$\hat{\gamma}_p$, $\hat{\gamma}_W$, $\hat{\gamma}_0$};
        \node[black] (s3) at (2.7,-3.5) {$\hat{\gamma}_0$, $\hat{\gamma}_W$, $\hat{\gamma}_p$};
        \node[black] (s3) at (-2.55,-3.5) {$\hat{\gamma}_p$, $\hat{\gamma}_0$, $\hat{\gamma}_W$};
        \node[anchor=west,black] (s4) at (0.55,-2.3) {$\hat{\gamma}_0$, $\hat{\gamma}_p$, $\hat{\gamma}_W$};

        \node[black] (s5) at (0.15,-4.1) {};
        \draw [->] (s4) -- (s5);



        \node[anchor=east] (s6) at (-4.2, 5.0) {\footnotesize $\nu_p$};
        \node[anchor=east] (s6) at (-4.2, -3) {\footnotesize $\frac{1}{n-3}$};
        \node[anchor=east] (s6) at (-4.2, -4.1) {\footnotesize $0$};

        \node[anchor=north] (s6) at (-4.1, -4.2) {\footnotesize $1$};
        \node[anchor=north] (s6) at (0.2, -4.1) {\footnotesize $\nu_p (n-3)$};
        \node[anchor=north] (s6) at (5.0, -4.1) {\footnotesize $n-3$};

        \node[anchor=south,rotate=90] (s6) at (-5.2, 0.5) {Composite covariate explained variance ($\nu_W$)};
        \node[anchor=north] (s6) at (0, -5.0) {Number of covariates ($p$)};

    \end{tikzpicture}
    \caption{Pairwise comparisons of relative efficiency of the treatment effect estimator between models with no covariate $\hat{\gamma}_0$, $p$ covariates $\hat{\gamma}_p$, and a composite covariate $\hat{\gamma}_W$ with respect to values $\nu_W$ and $p$. For each configuration, models are described by decreasing efficiency.
    $n$ = sample size, $\nu_p$ = variance explained by the $p$ covariates, $\nu_W$ = variance explained by the composite covariate.
    \label{fig:first}}
\end{figure}

To summarize, the relative efficiency of models (a) with no covariate, (b) with $p$ covariates, and (c) with the composite covariate $W$, can be compared.
Figure~\ref{fig:first} displays range values for $\nu_W$ and $p$ according to pairwise comparisons.
Due to the linearity of the generative model defined in Section~\ref{sec:model}, $\nu_W$ is upper-bounded by $\nu_p$.
As such, the y-axis, representing possible values of $\nu_W$, ranges between $0$ and $\nu_p$.
However, in practice, $\nu_p$ is not constant but monotonically increasing with $p$.
On the x-axis, the number of covariates, $p$, takes values between $1$ and $n-3$.
Firstly, from Equation~\eqref{eq:gainnup}, $\nu_W$ should be at least $1/(n-3)$ for $\hat{\gamma}_W$ to be as efficient as $\hat{\gamma}_0$.
This is represented by the horizontal line.
Secondly, Equation~\eqref{eq:nuWnup} induces that $\nu_W$ should be larger than $1-(1-\nu_p)(n-4)/(n-p-2)$ for the composite covariate to be more efficient than $p$ covariates.
This bound is represented by the curve starting at the top left corner.
Above the curve, $\hat{\gamma}_W$ is more efficient and below, $\hat{\gamma}_p$ is more efficient.
Thirdly, from Equation~\eqref{eq:gainnup}, when $p$ is larger than $\nu_p(n-3)$, $\hat{\gamma}_p$ is less efficient than $\hat{\gamma}_0$.
This threshold is represented by the vertical line Figure~\ref{fig:first}.

The three lines cross each other at the same point, hence defining six sets of values for $\nu_p$ and $p$ according the three pairwise comparisons.
In Figure~\ref{fig:first}, each set is identified by a unique ordering of the three estimators from most to least efficient.
These results show that $\hat{\gamma}_W$ becomes the most efficient estimators when $p$ increases.
In particular, a composite covariate does not need to be perfect and might be the best option even if $\nu_W \ll \nu_p$.

\bigskip
Replacing the covariates by a composite covariate $W$ estimated from previous data offers several advantages.
First, it improves the precision while limiting the loss in degrees of freedom.
The composite covariate approach could be seen as a way to borrow degrees of freedom from previous data.
Another advantage is to free the estimation of the treatment effect from modeling the covariates.
As such, one could use a non-linear model or machine-learning to estimate the composite covariate (\cite{Rasmussen2006,Hastie2009,Bishop2006}).
The explained variance of a non-linear composite covariate, $\nu_W$, is not upper-bounded by $\nu_p$ anymore.
Furthermore, the size of the composite covariate is only limited by previous data and $p$ could be larger than $n$.

Of note, composite covariates are already used in practice, e.g. through prognostic indexes~(\cite{Moons2009,INHLPFP93,Galea1992}).
A common non-linear example is the Body Mass Index (BMI)~(\cite{Keys1972}).
However, in this paper, we propose to use them as a way to optimize the precision of the estimated treatment effect.

\section{Simulation studies}
\label{sec:simulation}

To further illustrate the impact and relative gain of covariates or of a composite covariate on treatment precision and power, numerical simulations are performed.
All the simulations are performed with \texttt{R} software and are available in the supplementary materials.
The simulated studies are generated according to the model described in Section \ref{sec:model} (Equation \ref{eq:model0}).
The covariates $X$ and the random errors $\varepsilon$ are generated with independent Normal distributions.
The vector $\beta = (\beta_1, \beta_2, \dots)^T$ of true covariates weights is defined as:
\begin{align}
    \beta_k = 1 - \frac{1}{1 + \exp(- (k - 15) / 2)}, \quad \forall k \in \mathbb{N}.
\end{align}
This arbitrary choice is made to be representative of a common study setting where a few covariates explain most of the variance.
The treatment effect, $\gamma$, is computed for the studies to have 80\% of power without any covariate.
Previous data are also generated using exactly the same procedure.

\begin{figure}[ht]
    \center
    \begin{tikzpicture}
        \node (fig1) at (0,0) {\includegraphics[scale=0.60,page=1]{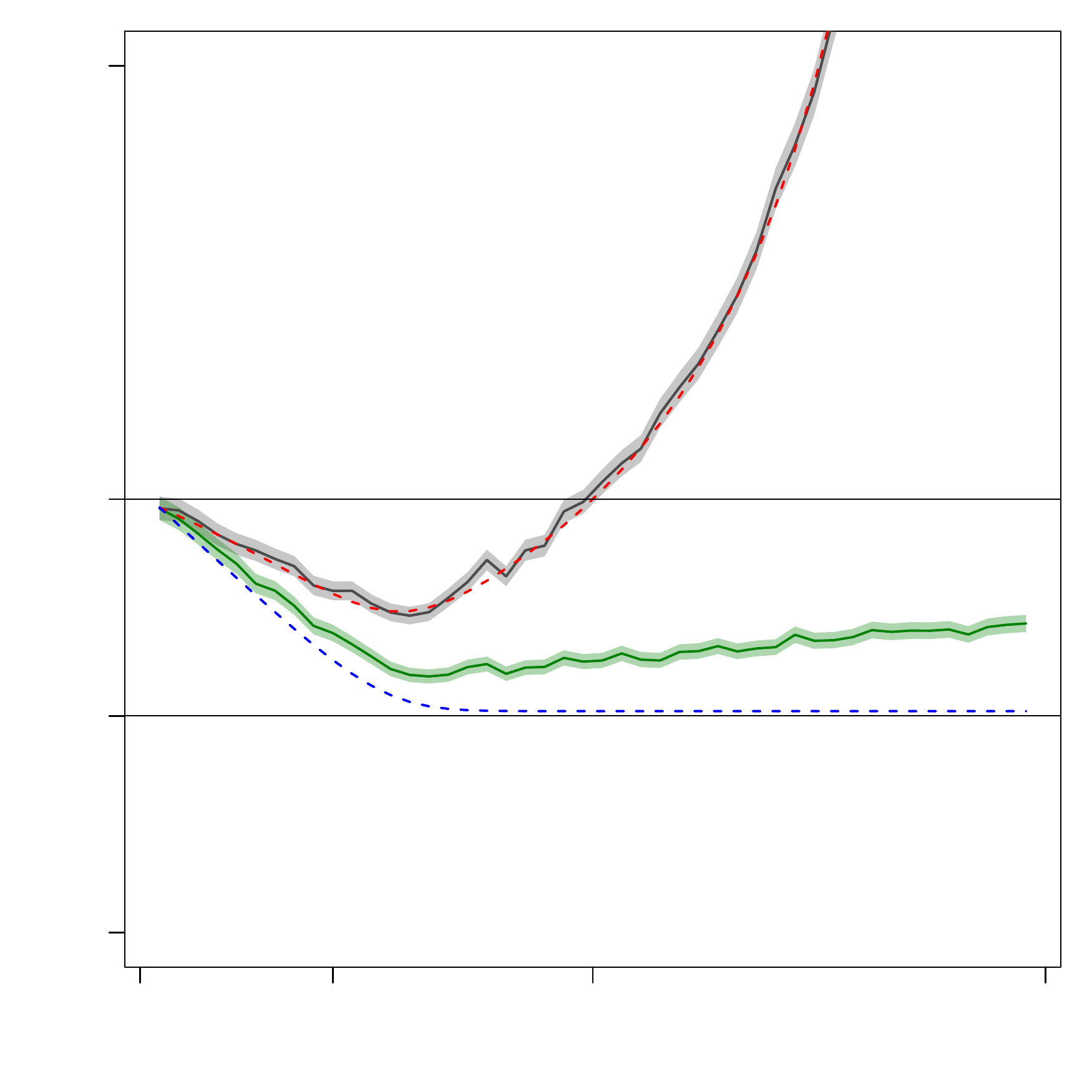}};
        \node[black] (s1) at (0.8,1.3) {(a)};
        \node[black] (s3) at (1.65,-1.35) {(b)};

        \node[anchor=east] (s6) at (-4.2, 4.7) {\footnotesize $2$};
        \node[anchor=east] (s6) at (-4.2, 0.5) {\footnotesize $1$};
        \node[anchor=east] (s6) at (-4.1, -1.7) {\footnotesize $1 - \nu_p$};
        \node[anchor=east] (s6) at (-4.2, -3.75) {\footnotesize $0$};

        \node[anchor=north] (s6) at (-4.0, -4.2) {\footnotesize $1$};
        \node[anchor=north] (s6) at (-2.05, -4.2) {\footnotesize $10$};
        \node[anchor=north] (s6) at (0.35, -4.1) {\footnotesize $\nu_p (n-3)$};
        \node[anchor=north] (s6) at (4.9, -4.1) {\footnotesize $n-3$};

        \node[anchor=south,rotate=90] (s6) at (-5.2, 0.5) {Relative efficiency};
        \node[anchor=north] (s6) at (0, -5.0) {Number of covariates ($p$)};

    \end{tikzpicture}
    \caption{Relative efficiency with respect to the estimation of the treatment effect without covariates.
    (a) The solid curve is the mean relative efficiency of the $p$ covariates and its 95\% confidence interval.
    The dashed curve is the expected value of this relative efficiency $(n - 3)(1 - \nu_p)/(n - 3 - p)$.
    (b) The solid curve is the mean relative efficiency of the composite covariate and its 95\% confidence interval.
    The dashed curve is the expected relative efficiency of an ideal composite covariate assuming $\nu_W = \nu_p$.
    $n$ = sample size, $\nu_p$ = variance explained by the $p$ covariates, $\nu_W$ = variance explained by the composite covariate.
    \label{fig:Relative1}}
\end{figure}

\bigskip
\noindent For the simulations, we choose to fix the total number of patients to $n=50$ ($25$ for each group) while varying the number of covariates, $p$, included in the estimation of the treatment effect.
Both $n$ and $p$ are directly linked to the degrees of freedom.
There is little interest in changing both parameters at the same time.
The number of patients in the previous data is set to $m = 100$.
The total amount of variance explained by all possible covariates, $\nu_\infty$, was arbitrarily set to $0.5$.
\begin{align}
    \nu_\infty = \lim_{p \to \infty} \nu_p = 0.5
\end{align}
Following the current simulation hypotheses, the variance explained by the $p$ first covariates, $\nu_p$, is:
\begin{align}
    \nu_p &= \nu_\infty \frac{\sum_{k=1}^{p} \beta_k^2}{\sum_{k=1}^{\infty} \beta_k^2}, \quad \forall p \in \mathbb{N}.
\end{align}
Using this result in Equation \eqref{eq:rep} gives the relative efficiency of the estimator $\hat{\gamma}_p$ with respect to $\hat{\gamma}_0$: using $p$ covariates for the current simulations as compared to not using them.
This relative efficiency is presented in Figure~\ref{fig:Relative1} with the dashed curve (a).
The associated solid curve is the estimated relative efficiency of the $p$ covariates with its 95\% confidence interval based on 10,000 simulations.
The estimated relative efficiency is computed as the ratio of the empirical variances of $\hat{\gamma}_p$ and $\hat{\gamma}_0$.
As defined in \cite{Morris2019}, the empirical variance is simply the estimated variance of $\hat{\gamma}$ over the simulations.

To illustrate the use of a composite covariate, a ridge regression is trained on the historical data for each simulation while changing the number of covariates (\cite{Hoerl1970}).
These ridge models are then used to predict the composite covariate, $W$, on the simulated studies.
The estimated relative efficiency of using $W$ ($\hat{\gamma}_W$ vs $\hat{\gamma}_0$) is presented with the solid line (b) in Figure~\ref{fig:Relative1} with its confidence interval.
As we can see on the figure, the use of a composite covariate can lead to an important gain in precision.
Of course, the gain depends on $\nu_W$, the variance explained by the composite covariate.
Similarly as for $p$ covariates, we can estimate the relative efficiency of a composite covariate approach assuming that $\nu_W = \nu_p$.
The relative efficiency of this ideal composite covariate is depicted in the figure with the dashed line (b).
The larger the amount of historical data, the closer the composite covariate is from this upper-bound.

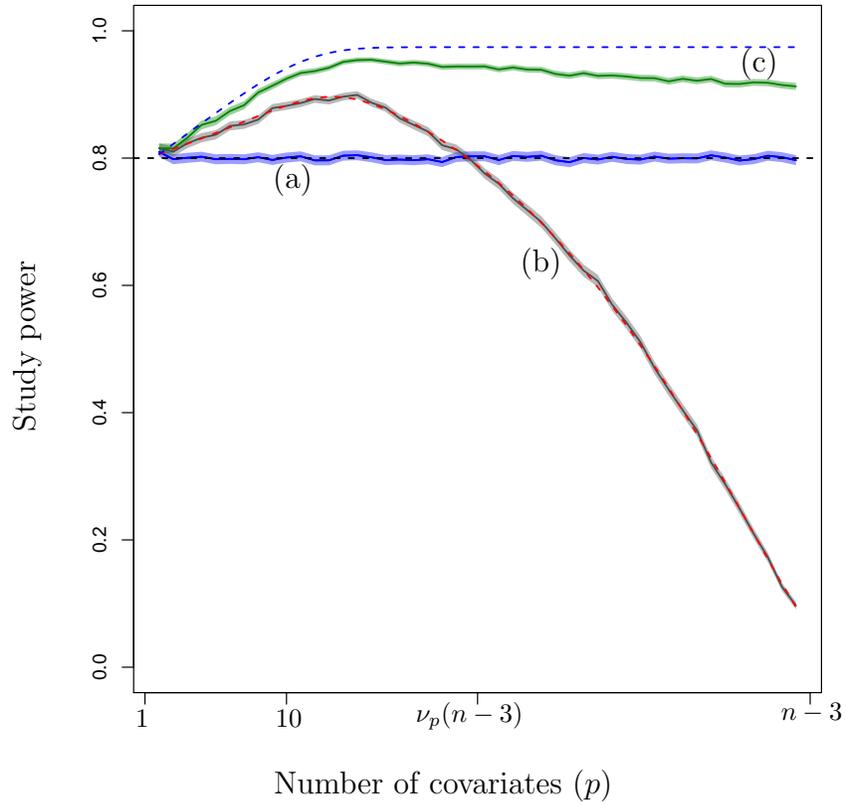
\begin{figure}[ht]
    \center
    \begin{tikzpicture}
        \node (fig1) at (0,0) {\includegraphics[scale=0.60,page=1]{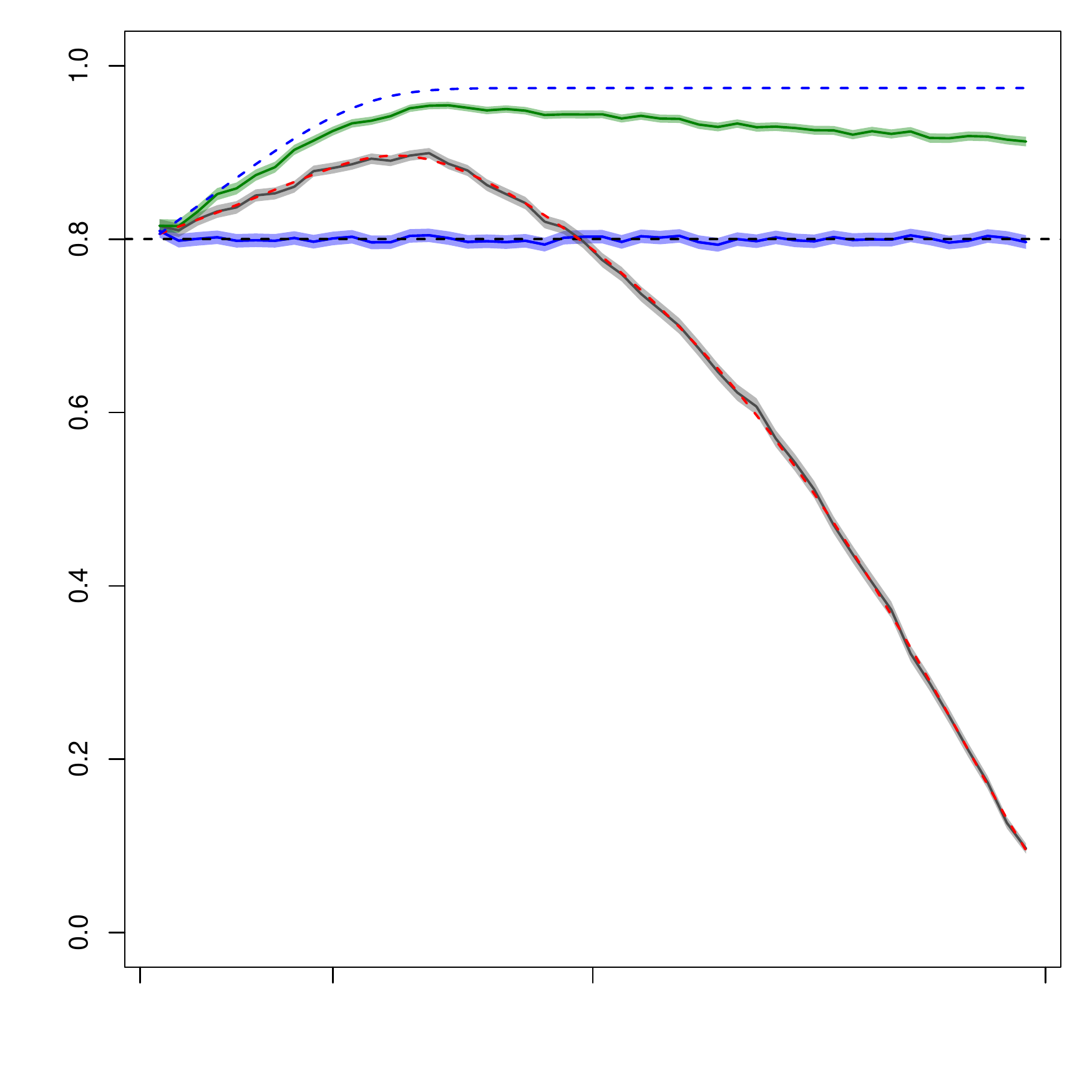}};
        \node[black] (s1) at (-2.0,2.7) {(a)};
        \node[black] (s3) at (1.3,1.6) {(b)};
        \node[black] (s3) at (4.2,4.25) {(c)};

        \node[anchor=north] (s6) at (-4.0, -4.2) {\footnotesize $1$};
        \node[anchor=north] (s6) at (-2.05, -4.2) {\footnotesize $10$};
        \node[anchor=north] (s6) at (0.35, -4.1) {\footnotesize $\nu_p (n-3)$};
        \node[anchor=north] (s6) at (4.9, -4.1) {\footnotesize $n-3$};

        \node[anchor=south,rotate=90] (s6) at (-5.2, 0.5) {Study power};
        \node[anchor=north] (s6) at (0, -5.0) {Number of covariates ($p$)};

    \end{tikzpicture}
    \caption{Study power of the three approaches and their 95\% confidence intervals: (a) without covariate , (b) with p covariates, and (c) the composite covariate.
    The dashed curves are the expected power (a) without covariate, (b) with p covariates and (c) with an ideal composite covariate.
    $n$ = sample size, $\nu_p$ = variance explained by the $p$ covariates.
    \label{fig:Power1}}
\end{figure}

\bigskip
\indent
Figure~\ref{fig:Power1} presents the power associated with the three approaches in the simulated studies: (a) without any covariate, (b) with $p$ covariates, and (c) with a composite covariate.
Without any covariate, the power is around 80\% as designed from the simulation protocol.
The use of the $p$ covariates brings a boost in power and then decreases (solid line (b)).
The solid line (c) shows the power gained by using the composite covariate.
The composite covariate power remains high even when $p$ increases.

The dashed line (b) represents the expected power of using the covariates with respect to the simulation hypotheses.
The dashed line (c) represents the expected power of an ideal composite covariate.
These power estimations are performed using the approach proposed by \cite{Shieh2020}.
Similarly, as for the relative efficiency, the advantage of the composite covariate grows with $p$.

\clearpage

\section{Discussion and conclusion}
\label{sec:conclusion}

Assessing correctly the treatment efficacy is of critical importance in randomized clinical trials.
However, since it is not ethical to expose too many patients to an unproven treatment, the sample size and power of phase I/II trials are often limited.
In this context, several statistical approaches have been developed to maximize the study power and the statistical precision of the treatment effect estimate.
One of such approaches, the analysis of covariance (ANCOVA), relies on baseline covariates to adjust for possible imbalance between study groups and to explain the variability of the patient's response improving the study power.

\bigskip
Including covariates associated with the study outcome could greatly improve the efficiency and power of the trial.
However, adding covariates in the analysis comes with a cost in degrees of freedom.
As such, regression adjustment should be seen as a trade-off between explained variance and loss of degrees of freedom.
There are many rules-of-thumb on the number of covariates that can be included in an analysis.
To the best of our knowledge, none of them balances both explained variance and degrees of freedom.

\bigskip
In this paper, we answered the question of the number of covariates while focusing on the precision of the estimated treatment effect in an ANCOVA.
Our result for the maximum number of covariates is a simple closed-form formula, $p < (n-g-1) \nu_p$, combining the number of patients and groups with the variance explained by those covariates.
We also proposed a simple method relying on available data to estimate the optimal number of covariates.
This data-driven approach can easily be applied in practice to plan for future trials.

\bigskip
Assuming data of previous studies to be available, we showed how to further improve the study power by fitting the covariates weights a priori.
Similarly, a composite covariate is fitted on previous data and replaces the individual covariates in the treatment effect estimation.
The composite covariate approach is already used in practice, e.g. through prognostic indexes (see \cite{Moons2009,INHLPFP93,Galea1992}).
With this paper, we investigated the use composite covariates specifically to optimize the precision of the treatment effect estimation.
Using a composite covariate allows to trade some explained variance to avoid the loss in degrees of freedom.
The associated gain is particularly important when the sample size is small and the number of covariates is large.

\bigskip
Considering the recent advances in placebo effect characterization~(\cite{Horing2014,Pereira2016,Vachon-presseau2018}), the composite covariate approach could have a major impact on future RCTs by disentangling the placebo response from the actual treatment efficacy.
The placebo effect is a complex phenomenon, individual-dependent with components linked to the subject's demography, psychology, sociology and disease intensity.
The highly multivariate aspect of the placebo makes any adjustment difficult.
The composite covariate approach could be used in this context to control for this major confounding factor in RCTs.


\section*{Acknowledgments}
The authors are grateful for the valuable feedback and suggestions provided by Marc Buyse
which greatly improved this article.


\clearpage
\appendix

\section{Generalization to $g$ groups \label{sec:moreGroups}}


In this section, we generalize the previous results to $g$ groups.
The vector~$z$, the treatment variable, is now taking values in ${1, \dots, g}$.
We denote by $\mu$ the vector of all group intercepts.
As previously, we can compute the variance of the estimator of $\mu$, with and without the $p$ covariates.
We denote by $\hat{\mu}_0$ the estimator of $\mu$ when no covariate are used in the model and $\hat{\mu}_p$ when $p$ covariates are included in the regression.

\bigskip
\noindent When there is no covariate, the sampling variance-covariance matrix of $\hat{\mu}_0$ can be written as
\begin{align}
    \mbox{Var}(\hat{\mu}_0 | z) = \sigma_u^2 \bm{D}
\end{align}
where $\bm{D} = \mbox{Diag}(1 / n_1, \dots, 1 / n_g)$, $n_j$ is the jth group size.
We have $n = \sum_{j=1}^{g} n_j$.

\bigskip
\noindent When there are $p$ covariates, the sampling variance-covariance matrix of $\hat{\mu}_p$ is
\begin{align}
    \mbox{Var}(\hat{\mu}_p | z, \X) = \sigma^2_\varepsilon (\bm{D} + \bar{\bm X}^T S_{XX}^{-1} \bar{\bm X})
\end{align}
where
$x_i = (x_{i1}, \dots, x_{ip})^T$ is the covariate vector for patient $i$,
$\bar{\bm{X}}_j = \sum_{i|z_i=j} x_i / n_j$ is the mean vector for treatment $j$,
$\bar{\bm X} = (\bar{\bm{X}}_1, \dots , \bar{\bm{X}}_g)$, and
$S_{XX} = \sum_{j = 1}^{g} \sum_{i|z_i=j} (x_i - \bar{\bm{X}}_j)(x_i - \bar{\bm{X}}_j)^T$.
We assume here, without any loss of generality, the covariates to be centered, $\sum_{i=1}^n x_i = \bm{0}_p$.

\bigskip
The treatment effects are computed using a contrast matrix $\bm{C}$ of size $c \times g$ of full row rank, satisfying $\bm{C}\bm{1}_g = \bm{0}_c$.
The treatment effect is then a vector of size $c \times 1$:
\begin{align}
    \hat{\gamma} &= \bm{C} \hat{\mu}
\end{align}
Its sampling variance-covariance matrix is respectively
\begin{align}
    \mbox{Var}(\hat{\gamma}_0 | z) &= \sigma_u^2 \bm{C}\bm{D}\bm{C}^T\\
    \mbox{Var}(\hat{\gamma}_p | z, \X) &= \sigma^2_\varepsilon \bm{C}(\bm{D} + \bar{\bm X}^T S_{XX}^{-1} \bar{\bm X})\bm{C}^T
\end{align}
Variances of the marginal distributions for the individual entries of the $\hat{\gamma}$ vector are on the diagonal of the variance-covariance matrix.
As such the sampling variance of the estimator of the $k$th entry of the $\gamma$ vector is
\begin{align}
    \left[\mbox{Var}(\hat{\gamma}_0 | z)\right]_{kk} &= \sigma_u^2 \bm{C}_k\bm{D}\bm{C}_k^T  \qquad \mbox{or}\\
    \left[\mbox{Var}(\hat{\gamma}_p | z, \X)\right]_{kk} &= \sigma^2_\varepsilon \bm{C}_k(\bm{D} + \bar{\bm X}^T S_{XX}^{-1} \bar{\bm X})\bm{C}_k^T
\end{align}
where $\bm{C}_k$ is the $k$th row of $\bm{C}$.
The ratio of the sampling variance of the two estimator is
\begin{align}
    \frac{\left[\mbox{Var}(\hat{\gamma}_p | z, \X)\right]_{kk}}{\left[\mbox{Var}(\hat{\gamma}_0 | z)\right]_{kk}} &= \frac{\bm{C}_k(\bm{D} + \bar{\bm X}^T S_{XX}^{-1} \bar{\bm X})\bm{C}_k^T}{\bm{C}_k\bm{D}\bm{C}_k^T} \frac{\sigma^2_\varepsilon}{\sigma_u^2}
\end{align}
As previously, we assume the covariates have independent and identical normal distribution for each patient.
From \cite{Shieh2020}, we then have
\begin{align}
    \frac{\bm{C}_k(\bm{D} + \bar{\bm X}^T S_{XX}^{-1} \bar{\bm X})\bm{C}_k^T}{\bm{C}_k\bm{D}\bm{C}_k^T} &= \frac{1}{(1 - B)}
\end{align}
where $B \sim \BetaDist(p/2, (n - p - g + 1)/2)$.
The relative efficiency becomes
\begin{align}
    RE_p &= E\left[\frac{1}{(1 - B)}\right] \frac{\sigma^2_\varepsilon}{\sigma_u^2} \\
    &= \frac{(n - g - 1)}{(n - p - g - 1)} \frac{\sigma^2_\varepsilon}{\sigma_u^2} \\
    &= \frac{(n - g - 1)}{(n - p - g - 1)} (1-\nu_p)
\end{align}
As a consequence, the $p$ covariates included in the model improve the statistical precision of the estimators if $RE_p < 1$, i.e., if
\begin{equation}
\nu_p > \frac{p}{n - g - 1}.
\end{equation}
Equivalently, the number of variables to be included should satisfy the inequality
\begin{equation}
p < (n - g - 1) \nu_p.
\end{equation}









\bibliographystyle{apalike}

\bibliography{library}

\end{document}